\documentclass[aps,prb,twocolumn,showpacs]{revtex4-1}
\usepackage{graphicx} 
\usepackage{epstopdf}
\usepackage{bm}       
\usepackage{amsfonts}
\usepackage{amssymb}
\usepackage{amsmath}
\usepackage{hyperref}
\usepackage{times}

\begin{document}

\title{Graphene Nano-Ribbons: Major differences in  the fundamental gap 
as its length  is increased either in the zig-zag or the armchair directions}

\author{J.A. Verg\'es$^{1,2}$, G. Chiappe$^{2,3}$, E. Louis$^{2,3}$}
\affiliation{
$^1$Departamento de Teor\'{\i}a y Simulaci\'on de Materiales, Instituto de
Ciencia de Materiales de Madrid (CSIC),
Cantoblanco, 28049 Madrid, Spain.\\
$^2$Unidad Asociada del CSIC and
Instituto Universitario de Materiales, Universidad
de Alicante, San Vicente del Raspeig, 03690 Alicante, Spain.\\
$^3$Departamento de F\'{\i}sica Aplicada,  Universidad
de Alicante, San Vicente del Raspeig, 03690 Alicante, Spain.}
\date{\today}

\begin{abstract}
Controlling the forbidden gap of graphene nano-ribbons (GNR) is a major challenge that has to be
attained if this attractive material has to be used in micro- and nano-electronics. Using an unambiguous notation
\{m,n\}-GNR, where $m (n)$ is the number of six carbon rings in the arm-chair (zig-zag) directions, we investigate how varies 
the HOMO-LUMO gap when the size of the GNR is varied by increasing either  $m$ or $n$, while keeping the other variable fixed. 
It is shown that no matter whether charge- or spin-density-waves  solutions are considered,
the gap varies smoothly when $n$ is kept fixed whereas it oscillates when the opposite is done, posing serious difficulties
to the control of the  gap. 
It is argued that the origin of this behavior is the fact that excess or defect charges or magnetic moments are mostly localised at zig-zag edges.

\end{abstract}
\maketitle

\section{Introduction}
 
Major improvements in bottom-up tecnologies are allowing the fabrication of Graphene Nano-Ribbons (GNR) with
well-defined shape  and size \cite{WT16,TR16,WZ16,XL16,YL16,WC15,LG14,CO13,TS13,RC12,VC15,KA12,BP13}. 
This is opening the possibility   of controlling the forbidden band gap \cite{KA12,SC06a,SC06b} and, thus, 
 widen the range of technological applications of graphene \cite{CG09,DA11,LV07}. Measuring
the nanoribbon conductance and/or using Scanning Tunneling Spectroscopy (STS)
to determine the Local Density of States (LDOS) have allowed the researchers
obtaining valuable information on the electronic structure around the HOMO-LUMO gap. Results have been already published for  7-AGNR (see below for notation)
\cite{WT16,TS13,KA12,RC12} and 13-AGNR \cite{CO13}. 

Although most data were taken on ribbons adsorbed on (111) surface of fcc
metals, very specially on Au(111) \cite{KA12,RC12,CO13,TS13,LC14}, 
 recently, several auhors have been able to lift off the surface a single graphene nanoribbon \cite{KA12,BP13,WT16}
by controlled pulling of one of the ribbon's ends using a STM. This technique is being applied to a variety of studies of considerable interest \cite{TR16,XL16}: 
 i)  keeping one of the GNR's end attached to the tip the ribbon was characterized before and after lifting by imaging and spectroscopy
 \cite{KA12}, ii) a reliable  transfer process of the lifted layer has allowed the
investigation of the transistor performance of GNR \cite{BP13}, iii) more recently, transferring the GNR to a thin NaCl deposit onto a gold substrate
has allowed, according to the authors of Ref. [\onlinecite{WT16}], a reliable characterization of the electronic structure of the ribbon.
Several techniques have been developed to produce GNR  free of defects \cite{XL16}, albeit in most cases GNR are fabricated by  means of bottom-up techniques on metal (preferently gold) surfaces and with the help of STM \cite{WT16,TR16,XL16,CO13}. Recently  etching of larger pieces of graphene has also been utilized \cite{LC14}.
Alhough these techniques have, up to recently, only been  applied to fabricate ribbons with arm-chair edges and rather narrow in the zig-zag direction, several works have been published in the current year that, modifying the procedures used to fabricate GNR with arm-chair edges,
reports successful fabrication of ribbons with zig-zag edges \cite{WT16,TR16,LC14}. The strategy followed in those works consists of growing the ribbon not along the direction of the carbon--halogen bond, but at an angle of
either 30¡ or 90¡ to it. The method,  altnough not free of vacancies and kinks that distort the edges, has a reasonable reliability having allowed topological and spectroscopic studies \cite{LC14}.

In the present work we investigate how the forbidden gap (actually the HOMO-LUMO gap) varies  as a function of the nano-ribbon length for several widths and either zig-zag or arm-chair edges. Albeit in the latter case the gap varies smoothly with the ribbon's length, in the former it oscillates making far more difficult obtaining ribbons with a defined gap width.The reason for this harmful behavior is that either staggered magnetization or charges are mostly located in the zig-zag edges. Then, if increasing the ribbon length does not imply varying the number of zig-zag atoms one may expect a rather constant gap. The opposite should occur when length is increased along the zig-zag direction. Our numerical results confirm these conjectures. Although in our previous work we discarded spin polarised solutions arguing that, i) once the mono-determinantal approximation for the many-body wavefunction is abandoned it is likely that charge density wave (CDW) solutions become more favorable, and ii) no experimental evidence of spin-polarized edges has been found whatsoever \cite{WT16}, in this work spin density wave (SDW) solutions are also considered as  they have been investigated by many different theory groups.

\begin{figure}{}
\includegraphics[width=\columnwidth]{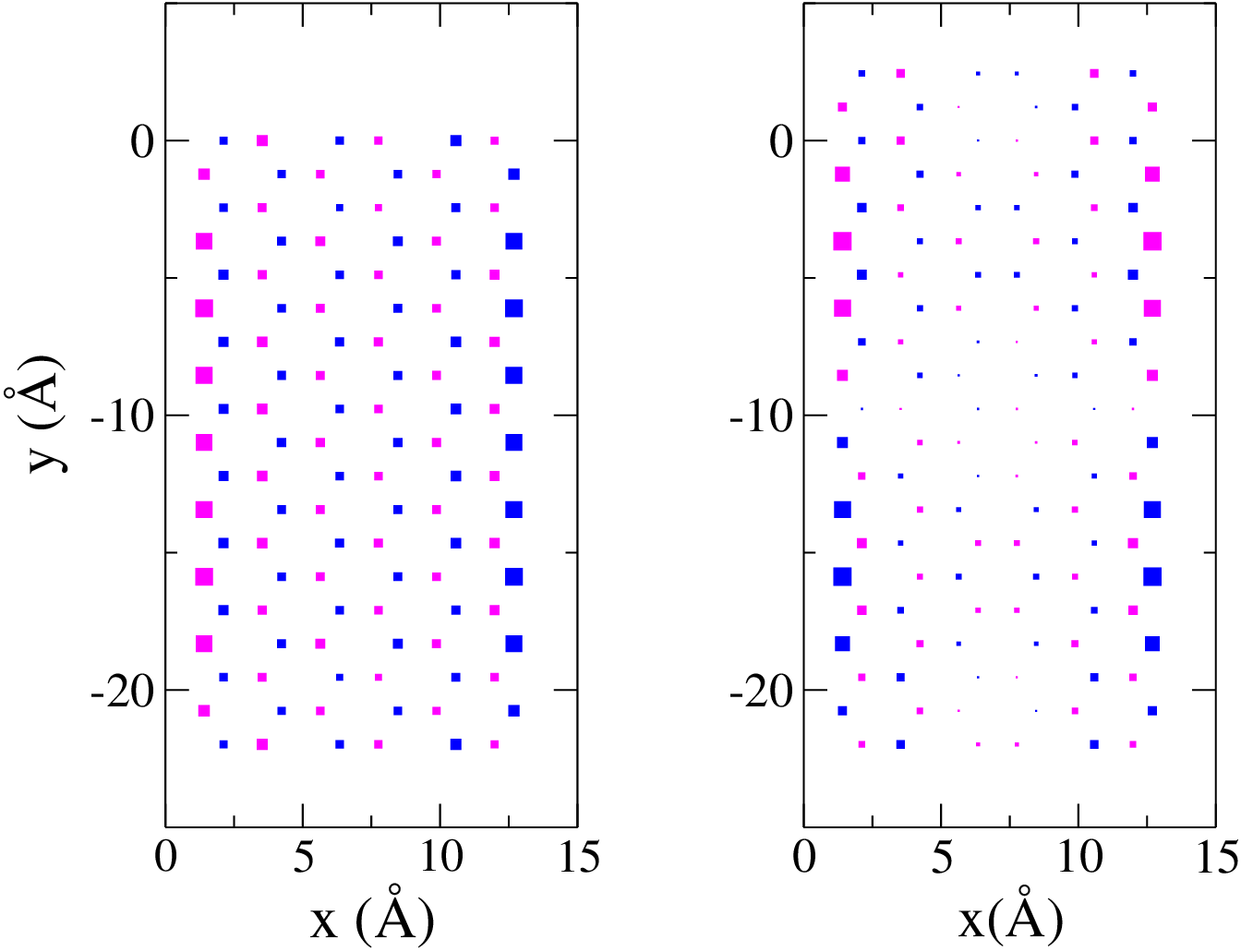}  \\
\caption{(Color online) Illustrates the notation used in this work to denote  finite Graphene-Nano-Ribbons (GNR). 
Those shown in the Figure are \{3,9\}-GNR (left) and \{3,10\}-GNR (right), repectively. In addition, each axis  has to be associated with a particular edge type; in
our case we choose the $x$ axis to correspond to  the armchair edge and the $y$ axis to the zig-zag edge.
In the more standard notation of Ref. [\onlinecite{SC06a}] the cluster shown in the left (right) panel  may either be 19-AGNR (21-AGNR) or 6-ZGNR 
(Z and A standing for zig-zag and armchair, respectively). The local $z$ component of the spin ($S_z$) of two SDW (spin density wave) solutions denoted
AF1-SDW (left) and SF0-SDW (right), see text, that differ in energy less than 0.002\% is also shown; Symbol size is proportional to actual  spin and color denotes up or down $S_z$. CDW solutions with similar arrangements of excess and defect charges are also found (see text). Maximum staggered magnetization 0.313.}
\label{(9-10)}
\end{figure}
\section{Hamiltonian, procedures and notation}
\subsection{The Pariser,Pople, Parr (PPP) Hamiltonian}
In this work we use the  model Hamiltonian  proposed by Pariser, Parr and Pople (PPP model)
\cite{PP53,Po53}, solved within the Unrestricted Hartree-Fock (UHF) approximation (see below), that
has been sucessfully applied to investigate the electronic structure of PAH \cite{CL15}.The PPP model includes both
local on-site and long-range Coulomb interactions. Only a single $\pi$ orbital per atom is considered.
The PPP Hamiltonian contains a non-interacting part $\hat H_{0}$ and a second term incorporating
electron-electron interactions $\hat H_{I}$:

\begin{equation}
{\hat H}  = {\hat H_{0}} + {\hat H_{I}} \;.
\label{eq:H}
\end{equation}

\noindent
Eventually, a core, constant term may be added to account for the
contribution of the rest of non-$\pi$ electrons to the total
energy\cite{VC09,VS09,VS10,SG11}. The non-interacting term is written as:

\begin{equation}
{\hat H_{0}}  = \epsilon_0 \sum_{i=1,N;\sigma}
{\hat c}^{\dagger}_{i\sigma} {\hat c}_{i\sigma} +
\sum_{<ij> ; \sigma}t_{ij} {\hat c}^{\dagger}_{i\sigma} {\hat c}_{j\sigma}\;,
\label{eq:H_0}
\end{equation}

\noindent
where the operator ${\hat c}^{\dagger}_{i\sigma}$ creates an electron
at site $i$ with spin $\sigma$, $\epsilon_0$ is the energy of the orbital, $N$
is the number of  orbitals
and $t_{ij}$ is the hopping between nearest neighbor pairs $<\!\!ij\!\!>$
(kinetic energy).

\begin{figure} {}
\includegraphics[width=\columnwidth]{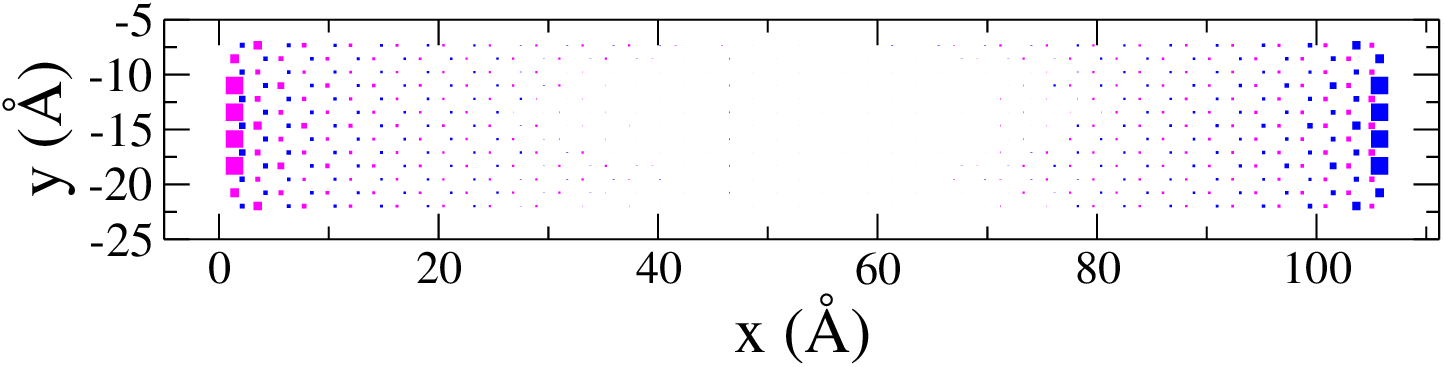} \\ 
\vspace{1.2cm}
\includegraphics[width=\columnwidth]{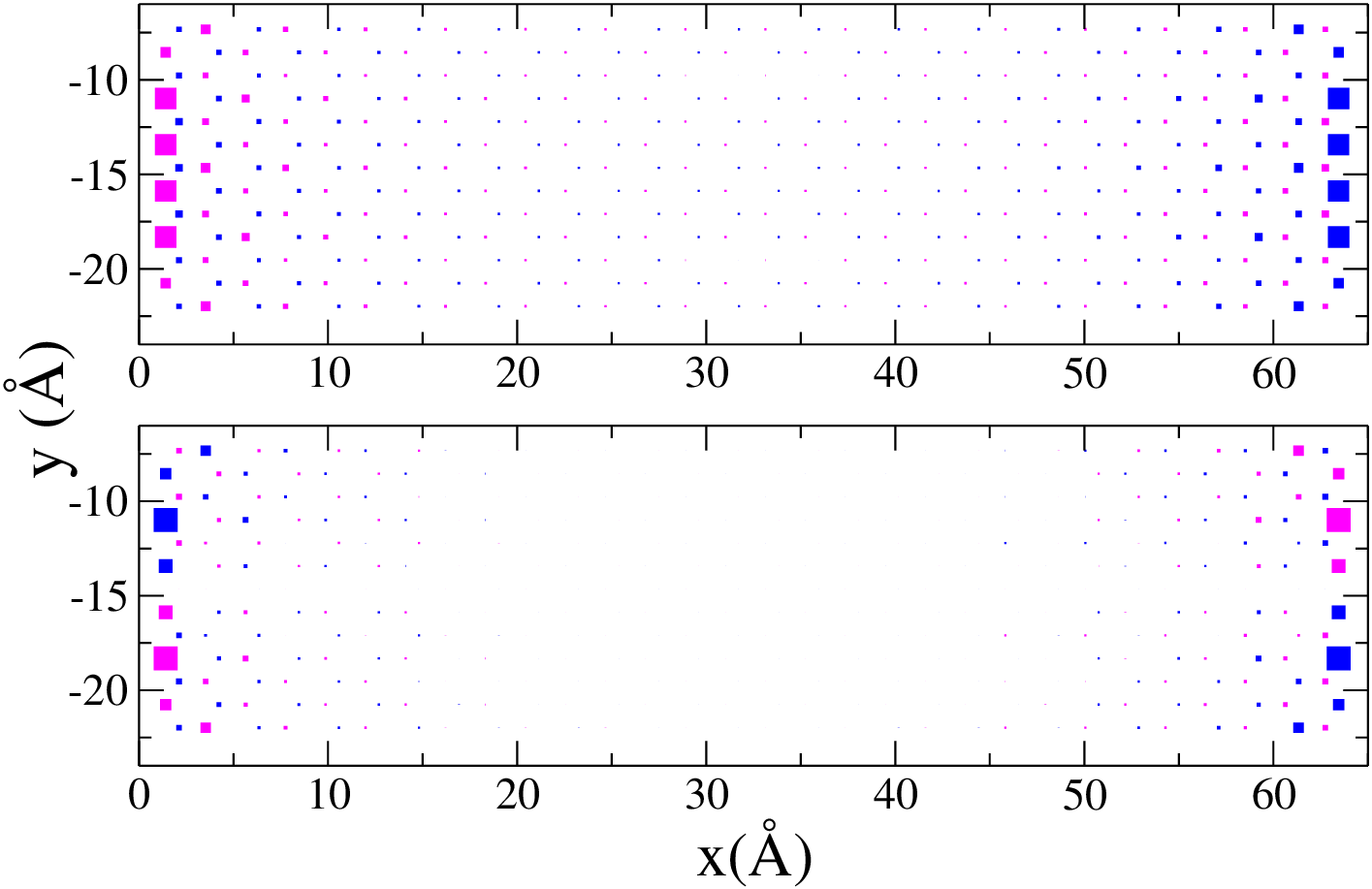}  \\
\caption{(Color online) Local distribution of the $z$-component of the spin $S_z$ in $\{15,6\}$ (two lower pictures) and 
$\{25,6\}$ (upper picture). Anti-symmetric solutions correspond to either A0-SDW (lower picture) or  A1-SDW (upper and middle pictures). Maximum staggered magnetization 0.35 (upper and center) and 0.52 (lower).}
\label{(15-25)}
\end{figure} 

In cases where the distance $d_{ij}$ between nearest neighbors pairs $<\!\!ij\!\!>$
significantly varies over the system, the hopping parameter may be scaled. For
instance in some PAH or even in defective graphene the C-C distance may differ
from its standard value $d_0 = 1.41$ \AA. In such cases one may use a scaling
adequate for $\pi$ orbitals \cite{Pa86}, namely

\begin{equation}
t_{ij}=\left(\frac{d_0}{d_{ij}}\right)^3t_0\;,
\end{equation}

\noindent where $t_0$ is a fitting parameter.
The assumption in using scaling laws is that the interatomic distance will
always be close to $d_0$, as occurs in most cases.

The interacting part is given by:

\begin{equation}
{\hat H_{I}}  =
U\sum_{i=1,N}{\hat n}_{i\uparrow} {\hat n}_{i\downarrow}+{\frac{1}{2}}
\sum_{i,j=1,N;i \neq j}V_{ij} ({\hat n}_i-1) ({\hat n}_j-1)\;,
\label{eq:H_I}
\end{equation}

\noindent
where $U$ is the on-site Coulomb repulsion,
$V_{ij}$ is the inter-site Coulomb repulsion and
the total electron density for site $i$ is

\begin{equation}
{\hat n}_i= {\hat n}_{i\uparrow} + {\hat n}_{i\downarrow}\;,
\end{equation}

In incorporating the Coulomb interaction $V_{ij}$ one may choose the unscreened
Coulomb interaction \cite{Ma00}, although it is a common practice the use of some
interpolating formula. In the case of PAH that proposed by Ohno\cite{Oh64}
has a wide acceptance,

\begin{equation}
V_{ij}=U\left[1+\left(\frac{U}{e^2/d_{ij}}\right)^2\right]^{-1/2}\;.
\end{equation}

\noindent
Using this interpolation scheme implies that no additional parameter is
introduced and, consequently,  $U$ remains as the single parameter associated to
interactions.

\subsection{Cluster Notation}
The notation used to identify graphene nano-ribbons (GNR) is illustrated in Figs. (\ref{(9-10)}) and (\ref{(15-25)}). Actually is identical to
that proposed in Ref. [\onlinecite{WC15}]. Each GNR is characterised by two indexes $\{m,n\}$ denoting the number of benzene rings in the armchair direction $m$ (or $x$ direction) and the zig-zag $n$ (or $y$ direction). This avoids possible ambiguities derived from an earlier and widely used notation which explicitly referred to arm-chair GNR ($N_z$-AGNR) or zig-zag GNR ($N_a$-ZGNR) ribbons \cite{SC06a}. It is useful connecting the present notation with the latter one proposed to name infinitely long ribbons (in infinitely long ribbons the ambiguities mentioned in the caption of Figs. (\ref{(9-10)}) do not longer show up. In a ribbon infinitely long in the arm-chair (zig-zag) direction $N_a=2n+1$ ($N_z=2m$), where $N_a$ ($N_z$)  the number of edge atoms in each case. The advantages of the notation used here are clearly illustrated in the  just mentioned Figures. 
 
\begin{figure}
\includegraphics[width=\columnwidth]{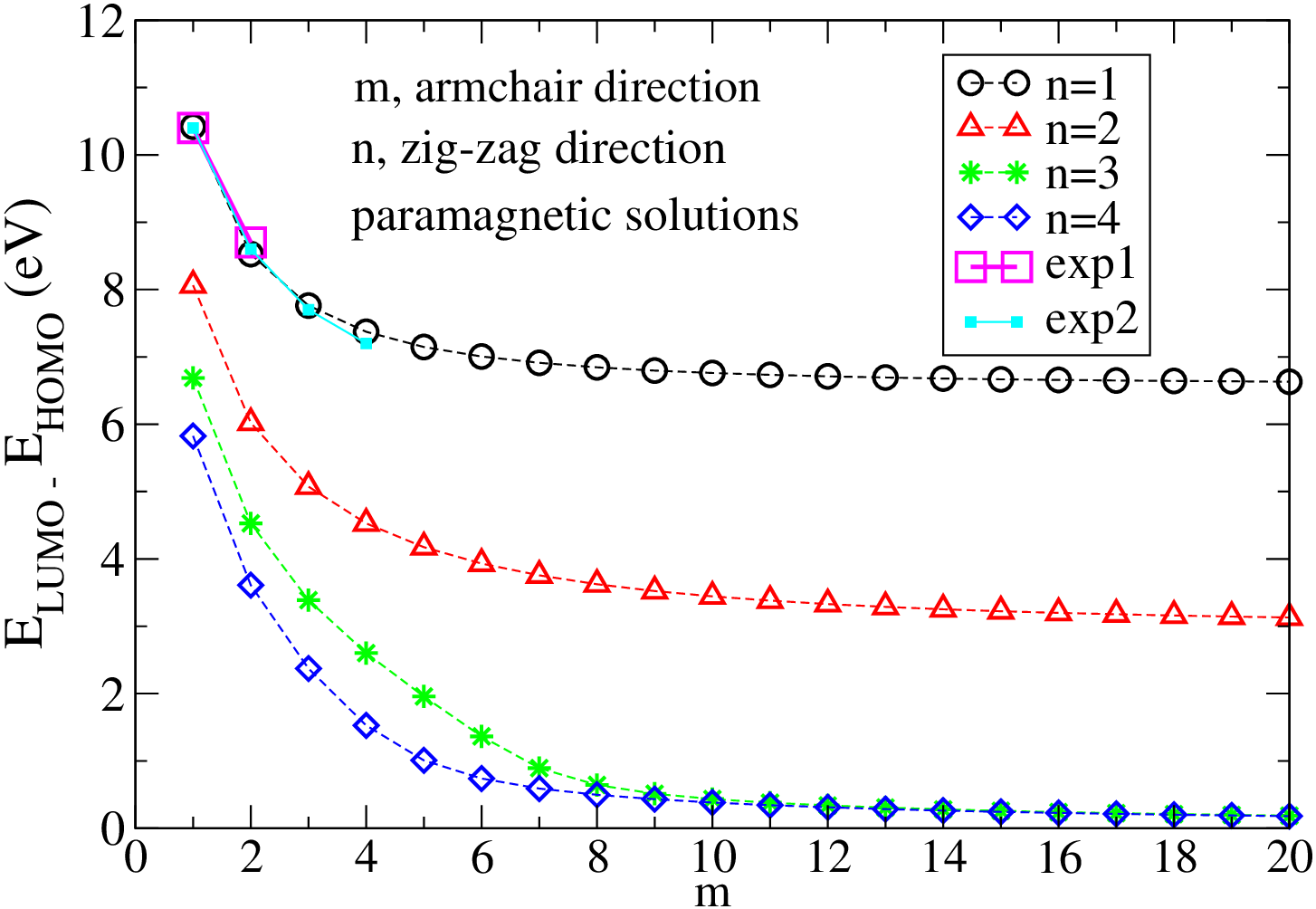}  \\
\vspace{1.2cm}
\includegraphics[width=\columnwidth]{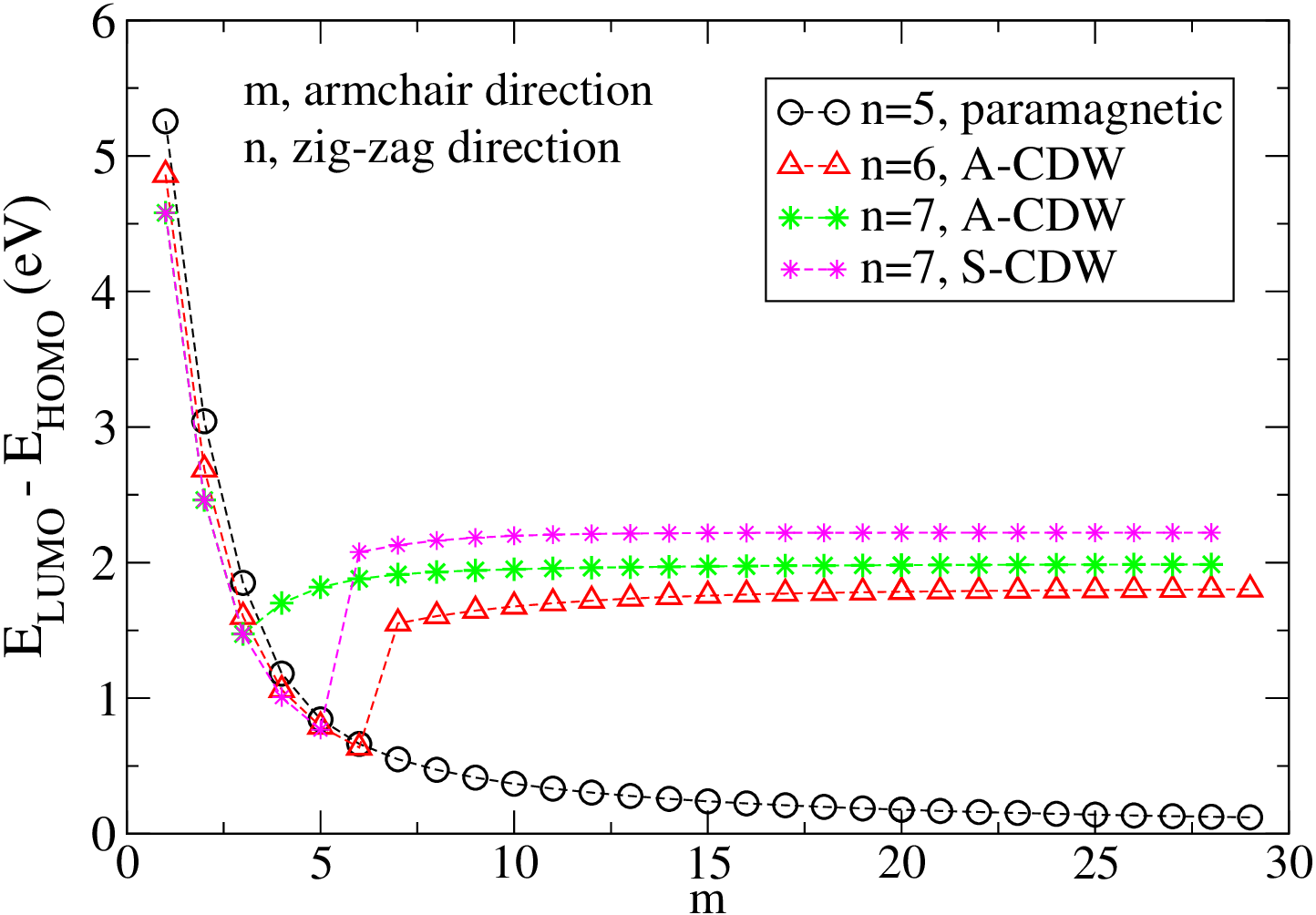} \\
\caption{(Color online) Calculated $E_{LUMO}-E_{HOMO}$ for paramagnetic solutions  in GNR of dimensions $\{1-20,1-4\}$ (upper)  and $\{1-28,5\}$ (lower)
and A-CDW solutions in GNR of dimensions $\{1-28,6-7\}$. In the case of $\{1-28,7\}$ resuts for both A-CDW and S-CDW solutions are shown.
Experimental results for GNR of dimensions $\{1-4,1\}$  [\onlinecite{SK84},\onlinecite{SS94},\onlinecite{MD83},\onlinecite{NA06},\onlinecite{RT02}]
are included in the upper plot (see also [\onlinecite{YL16}]). }
\label{GAP-CDW-m}
\end{figure}

\subsection{Procedures}

As noted above, in this work,
the PPP Hamiltonian has been solved within the UHF approximation. All
two operators terms (as given by Wick's theorem) conserving charge and spin have
been included:
The non-diagonal term plays an important role whenever crystalline periodicity
is absent. These includes
interaction between distant defects, presence of impurities or
surfaces, etc.\cite{SS10,CL15}.
The correct description of small finite systems like GNRs also requires
such a careful description.

The parameter values used hereafter have been derived from exact solutions of many spin
states of neutral and charged small molecules \cite{SG11,VC09,VS09,VS10}. They are:
$\epsilon_0$=-7.61 eV, $t_0$=-2.34 eV and $U$=8.29 eV.
Clusters containing up to about
1000 $\pi$--orbitals have been investigated in this work.
This has required the use of a simplified
treatment of interactions such as Hartree-Fock (HF). Nonetheless, we shall show
that even at such low level of approximation it is possible, in this case, to
attain results that throw light on experimental data and/or more sophisticated
theoretical calculations.
All ribbons here investigated had  the edge dangling $\sigma$ bonds passivated by hydrogen atoms.

\begin{figure}
\includegraphics[width=\columnwidth]{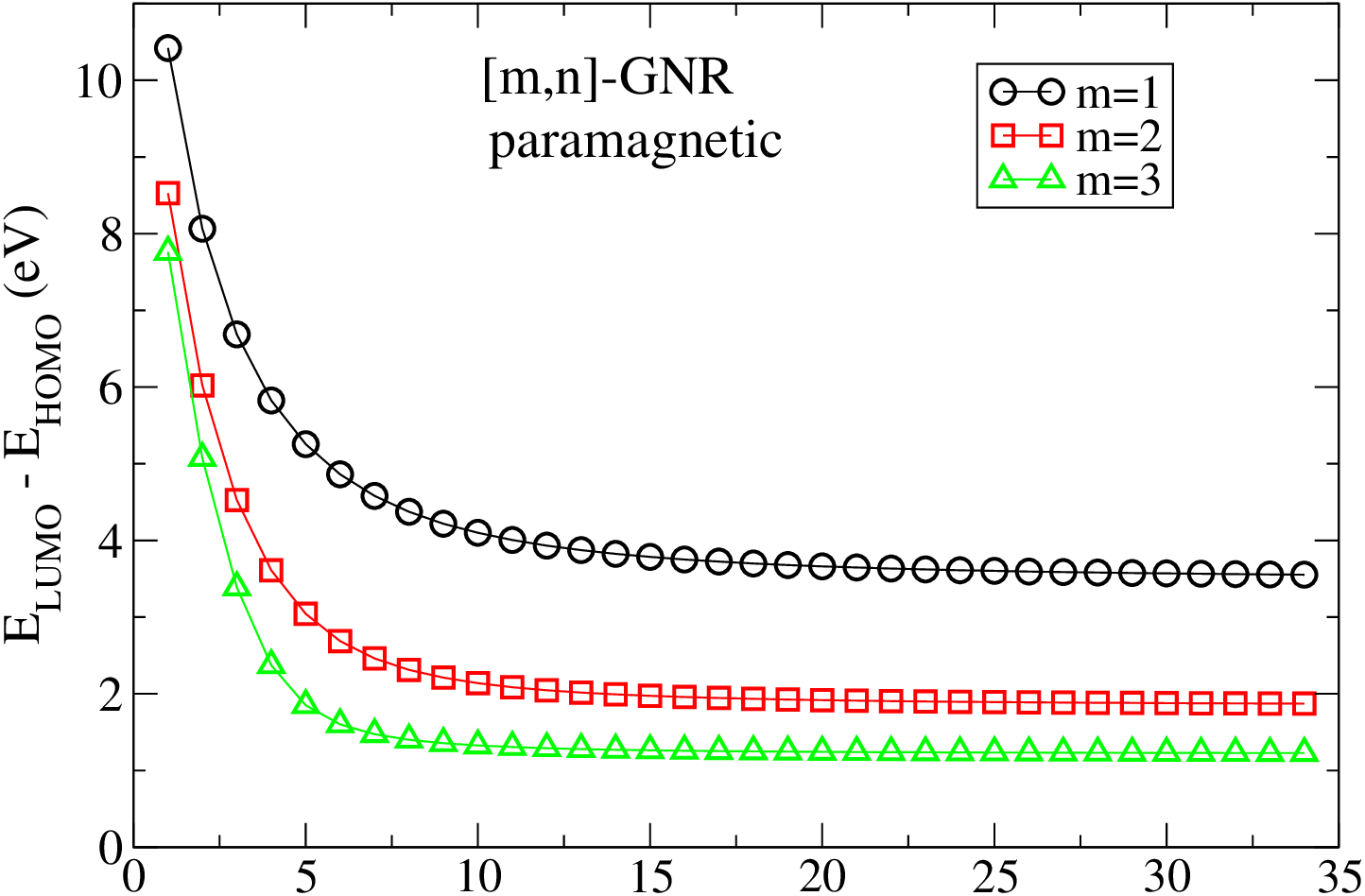}  \\
\vspace{1.2cm}
\includegraphics[width=\columnwidth]{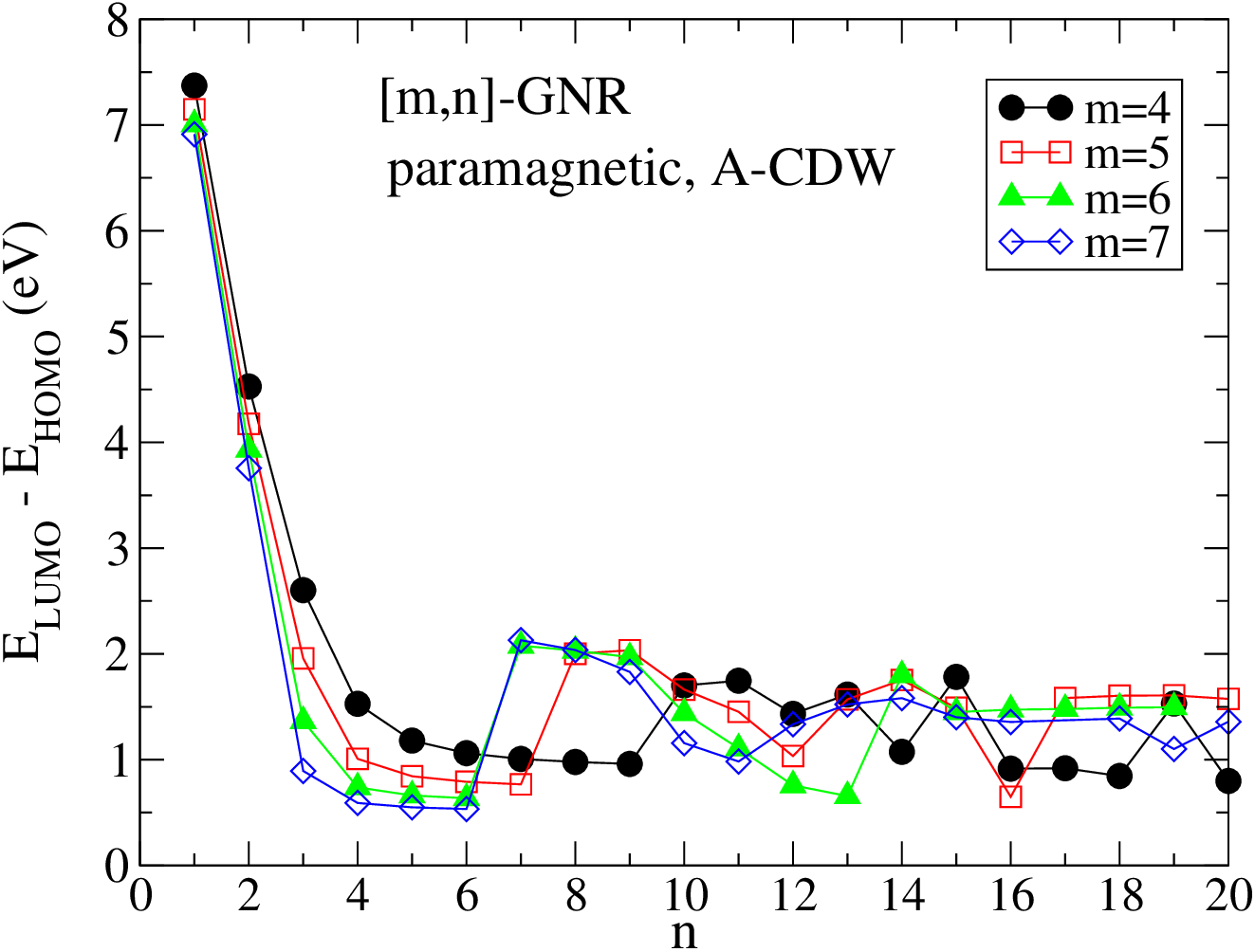} \\
\caption{(Color online)Calculated $E_{LUMO}-E_{HOMO}$ for paramagnetic solutions  in GNR of dimensions $\{1-3,1-34\}$ (upper)  and A-CDW solutions (for $n \leq $ 6-9, depending on $m$, the solution is paramagnetic) in $\{4-7,1-20\}$ GNR  (lower). }
\label{GAP-CDW-n}
\end{figure}

\begin{figure}
\includegraphics[width=\columnwidth]{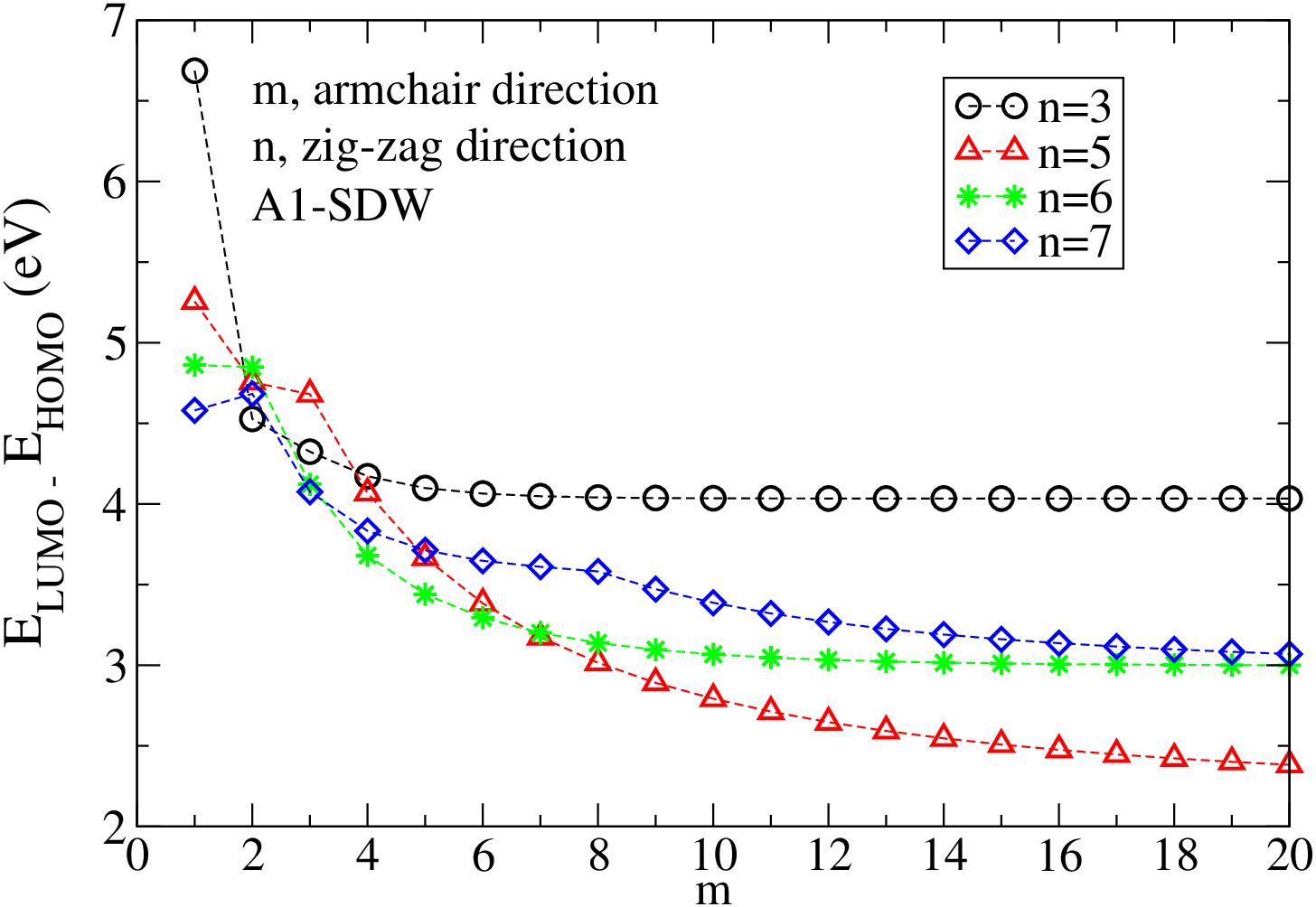}  \\
\vspace{1.2cm}
\includegraphics[width=\columnwidth]{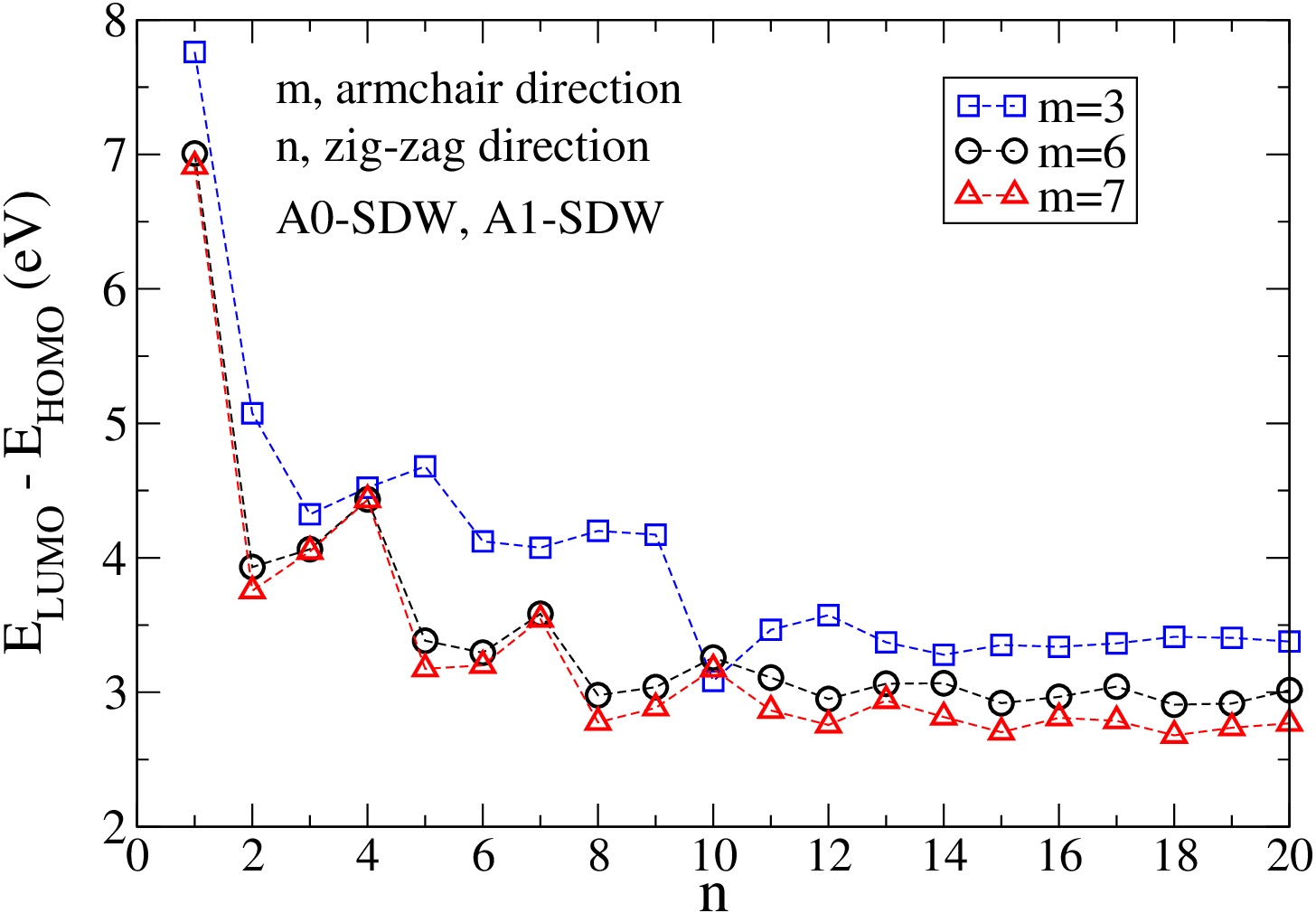} \\
\caption{(Color online) Calculated $E_{LUMO}-E_{HOMO}$ for antiferromagnetic solutions in GNR of dimensions $\{1-20,(3,5-7)\}$ (upper) and $\{(3,6,7),1-20\}$ (lower). In the first case (upper panel)  solutions are always of the A1-SDW type but for the smallest $m$ which show no magnetic polarization. In the second case, however, transitions between A0- and A1-SDW solutions occur as $n$ increases (these solutions differ  less than 0.02\% in energy). See also Fig. \ref{(9-10)}.}
\label{GAP-SDW-m-n}
\end{figure}
\vspace{2cm}

\subsection{Mean field solutions notation}
Three types of electronic configurations are obtained by solving UHF-PPP on GNR, namely, paramagnetic (P), Charge Density Waves (CDW) and Spin Density Waves (SDW). First note that the P configuration here obtained slightly differs from the strictly free electron solution ($U$ =0); this is due to
 the non-diagonal terms that the most general UHF approximation of the PPP model includes (see preceeding subsection). As regards charge and spin density waves solutions they can be classified as: S-CDW symmetric upon reflection through the $y$ axis passing by the center of the ribbon (no excess charge in both zig-zag edges). A-CDW anti-symmetric  upon reflection through the $y$ axis passing by the center of the ribbon (excess or defect charge at the two edges, although neutral the two halves of the GNR). A1-SDW anti-symmetric upon reflection through the $y$ axis passing by the center of the ribbon, positive $z$ component of the spin $S_z$ in one edge and negative at the opposite. A0-SDW anti-symmetric upon reflection through the $y$ axis passing by the center of the ribbon, but total $z$ component of the spin $S_z=0$ on both edges. The SDW solutions are illustrated in Figs. (\ref{(9-10)}) and (\ref{(15-25)}) of this work and the CDW solutions in Fig. 9 of Ref. [\onlinecite{VC15}], This notation will be used hereafter.

\section{Results}
Fig. \ref{GAP-CDW-m} shows the results for $\{m,1-7\}$-GNR. For $m\leq5$, the gap decreases monotonically and smoothly with the ribbon length (or, equivalently, $m$). Apparently, for $n$=1,2 (upper panel) the gap tends to a constant as $m$ increases, while it tends to zero for $n$=3-5. Beyond $n$=5 (lower panel), the CDW shows up \cite{VC15} for a length $m$ that decreases as $n$ increases. The emergence of the A-CDW implies an abrupt increase of the gap. Beyond this sharp increase the gap varies smoothly, almost remaining constant, with the ribbon length. As shown in the upper panel of the Figure, the calculated gap for $n$=1 is in excellent agreement with the available experimental data \cite{SK84,SS94,MD83,NA06,RT02}.  Finally, in the lower panel of the Figure, the results for the S-CDW solution in the ribbon with $n$=7 are also depicted. It should be noted that albeit  the gap in the symmetric solution is 0.24 eV higher than in the A-CDW solution, their energies differ in this case in less than 0.01\%, 
\begin{figure}
\includegraphics[width=\columnwidth]{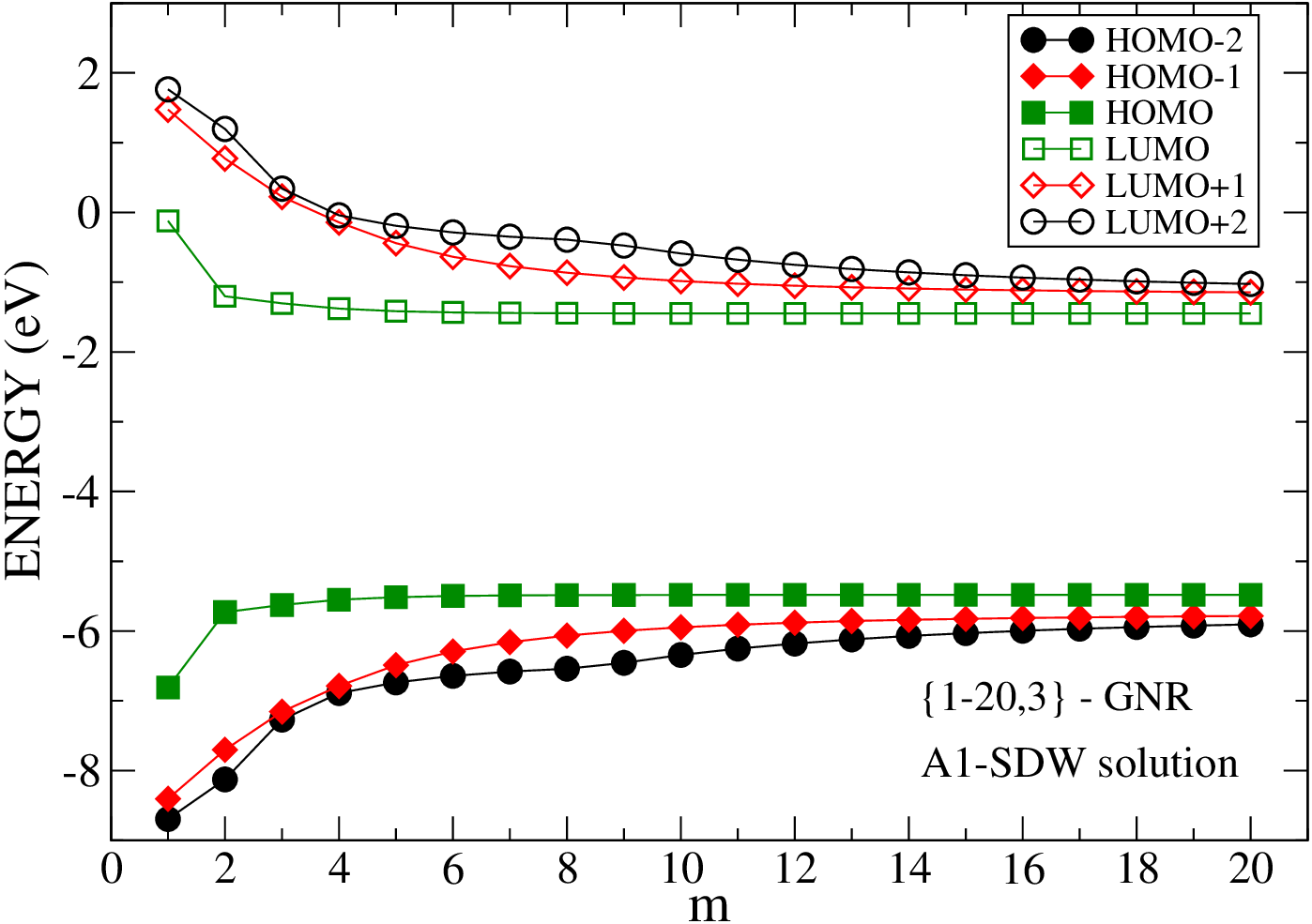}  \\
\vspace{1.2cm}
\includegraphics[width=\columnwidth]{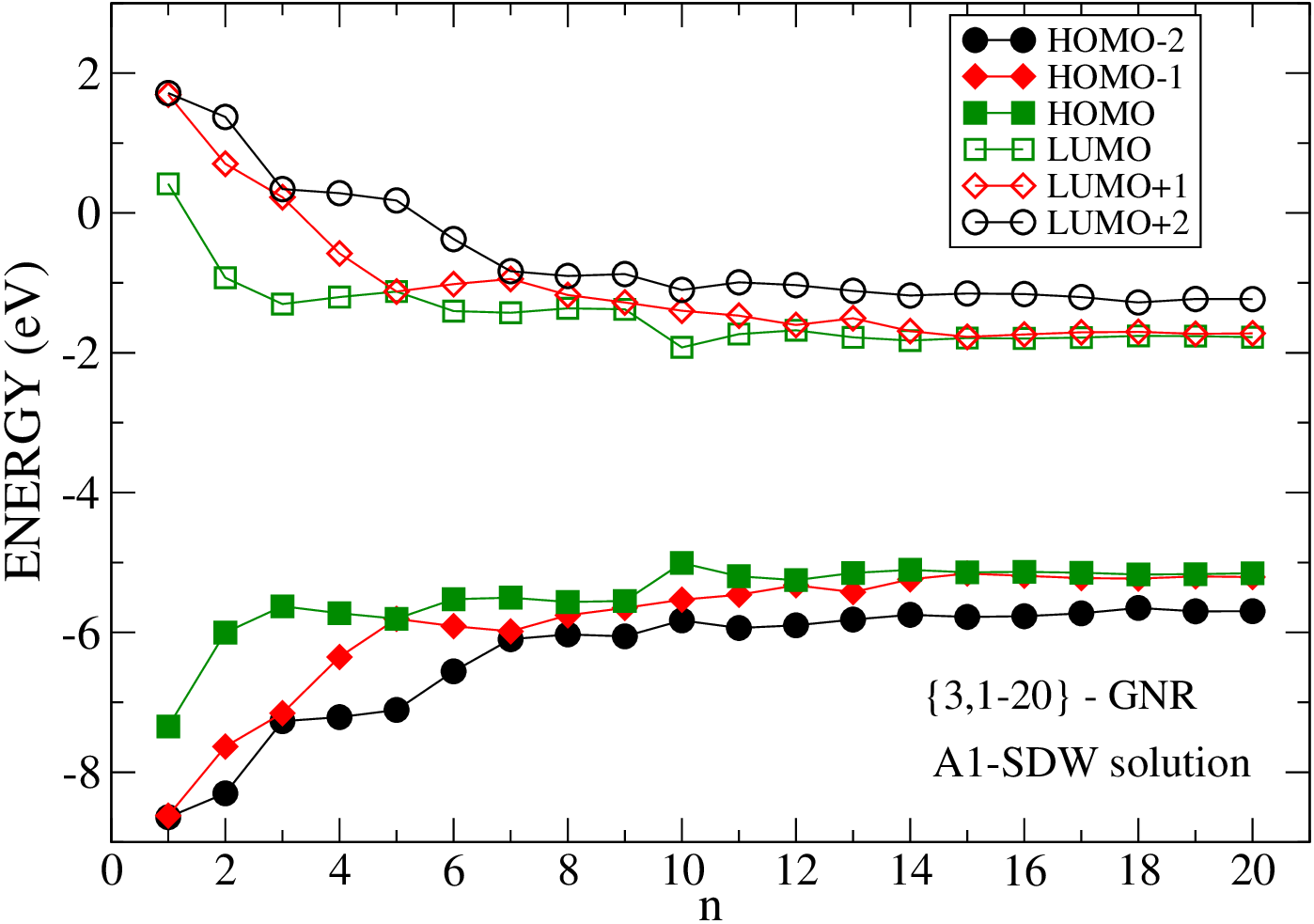}  \\
\caption{(Color online) Energies of the three upper HOMO and the three lower
LUMO versus the nanoribbon length in the armchair direction ($m$). The results correspond to A1-SDW
solutions of the PPP Hamiltonian on $\{1-20,6\}$ graphene nanoribbons (see Fig. 1 for notation).}
\label{E-SDW-3-m-n}
\end{figure}

It is pertinent comparing the present results with those reported by Yeh and Lee \cite{YL16} for the \{20,1\}-GNR, obtained with a variety of KS-DFT based methods. The results for the magnitude named by the authors $E_g(3)$, identical to that  plotted in Fig. \ref{GAP-CDW-m}, are depicted in Fig. 6c of Ref. 
 [\onlinecite{YL16}]. Only the results obtained from the $\omega$B97 and $\omega$B97X functionals do reasonably agree with the available
experimental data, although not as much as the results presented here as they are slightly higher.
 
Fig. \ref{GAP-CDW-n}  shows the results for $\{1-7,n\}$-GNR. In this case the ribbon is paramagnetic for $m$=1-3 and all values of $n$ shown in the Figure.
As in the previous case, the gaps decrease smoothly and monotonically with $n$. However,  it cannot be concluded whether  they will or they will not vanish for infinitely long ribbons. On the other hand, for $n > 3$ and a value of $m$ that decreases as $n$ increases, the charge density wave shows up. Beyond that point the gap oscillates in an unpredictable manner. These oscillations are surely due to the increase of the number of edge states with $n$ \cite{SC06a}.

We turn now to discuss the characteristics of the spin polarised solutions. Fig (\ref{GAP-SDW-m-n})  shows te gap in \{1-20,3-7\}-GNR and \{3-7,1-20\}-GNR. In the first case, whereby the ribbon grows in the armchair direction at constant width, the  gap varies smoothly with $m$ decreasing monotonically beyond $m > 2$, coinciding with the size at which the SDW replaces the paramagnetic solution.  In addition it varies in a manner consistent with the results reported in Ref. [\onlinecite{SC06a}], namely, the gap  of the three families that can be differentiated  goes as gap ($2n=3p$) $>$ gap($2n=3p-1$)$ >$ gap($2n=3p-1$), where $p$ is an integer. Moreover, within each family the gap decreases with the ribbon wdith, compare results for $n$=3  with those for $n$=6. This qualitative agreement (remind that ribbons in Ref. [\onlinecite{SC06a}] are infinite in one direction) is found for the LDA results reported in that work and not for those obtained with a one-electron tight-binding method which, among other features, predict a zero gap for the $3p+1$ family, no matter the ribbon width. Recent studies of $\{m,3\}$ \cite{WT16} with $m$ up to 48 provided the following results: i) experimental data (GW calculations) gave a gap between states localised at zig-zag edges of 1.8 (2.8) eV, a result  almost independent of ribbon length, and, ii) a delocalised (or bulk) states gap that did vary with the ribbon length being  around 2.8 eV for $m=48$. Our  calculations gave larger gaps around 4 (4.8) for localised (bulk) states that  did not vary (decrease) with the ribbon length.

It should be mentioned that all results shown in the Figure correspond to A1-SDW solutions (left panel in Fig. \ref{(9-10)} and upper and midle panels in Fig. \ref{(15-25)}). This solution shows, as discussed in Ref. [\onlinecite{SC06b}], half-metallicity. However, as already mentioned here,  there is  another SDW solution, referred to as A0-SDW, which has an energy very close to that of A1-SDW (only 0.002\% higher) that, having a null total $z$-component of the spin on the zig-zag edge atoms (see right panel in Fig. \ref{(9-10)} and lower panel in Fig. \ref{(15-25)}), cannot show  half-metallicity. 

Results for the gap in \{3-7,1-20\}-GNR are shown in the lower panel of Fig (\ref{GAP-SDW-m-n}). As in the previous case, the gap decreases with the ribbon length $n$, however a dramatic change is noted, namely, the  smooth decrease observed beyond $m > 2$ in\{1-20,3-7\}-GNR, is replaced by  oscillations as already observed in the case of CDW solutions. {\it Confirming this behavior experimentally will discard  this type of ribbon for technological applications} \cite{note}.

The  three HOMO having the highest energy and the three LUMO having the lower one are plotted in Fig. \ref{E-SDW-3-m-n}   versus the ribbon length, for  \{1-20,3\}-GNR and \{3,1-20\}-GNR. Again it is noted that while in the first case energy levels, and in particular the band gap, vary (actually decrease) smoothly with the ribbon length $m$, in the second case they oscillate in an irregular manner (in comparing results of Fig. \ref{E-SDW-3-m-n} with those of Fig. (\ref{GAP-SDW-m-n})  note that the  energy range in the former is twice that in the latter). Finally it is worth mentioning that while for $n > 15$ the two lower LUMO and two upper HOMO are almost degenerate, no degeneration is observed in ribbons elongated in the armchair direction \{1-20,3\}-GNR.
This is in agreement with results reported in Refs. [\onlinecite{NF96},\onlinecite{YP07}].

\begin{figure}{}
\vspace{1cm}
\includegraphics[width=\columnwidth]{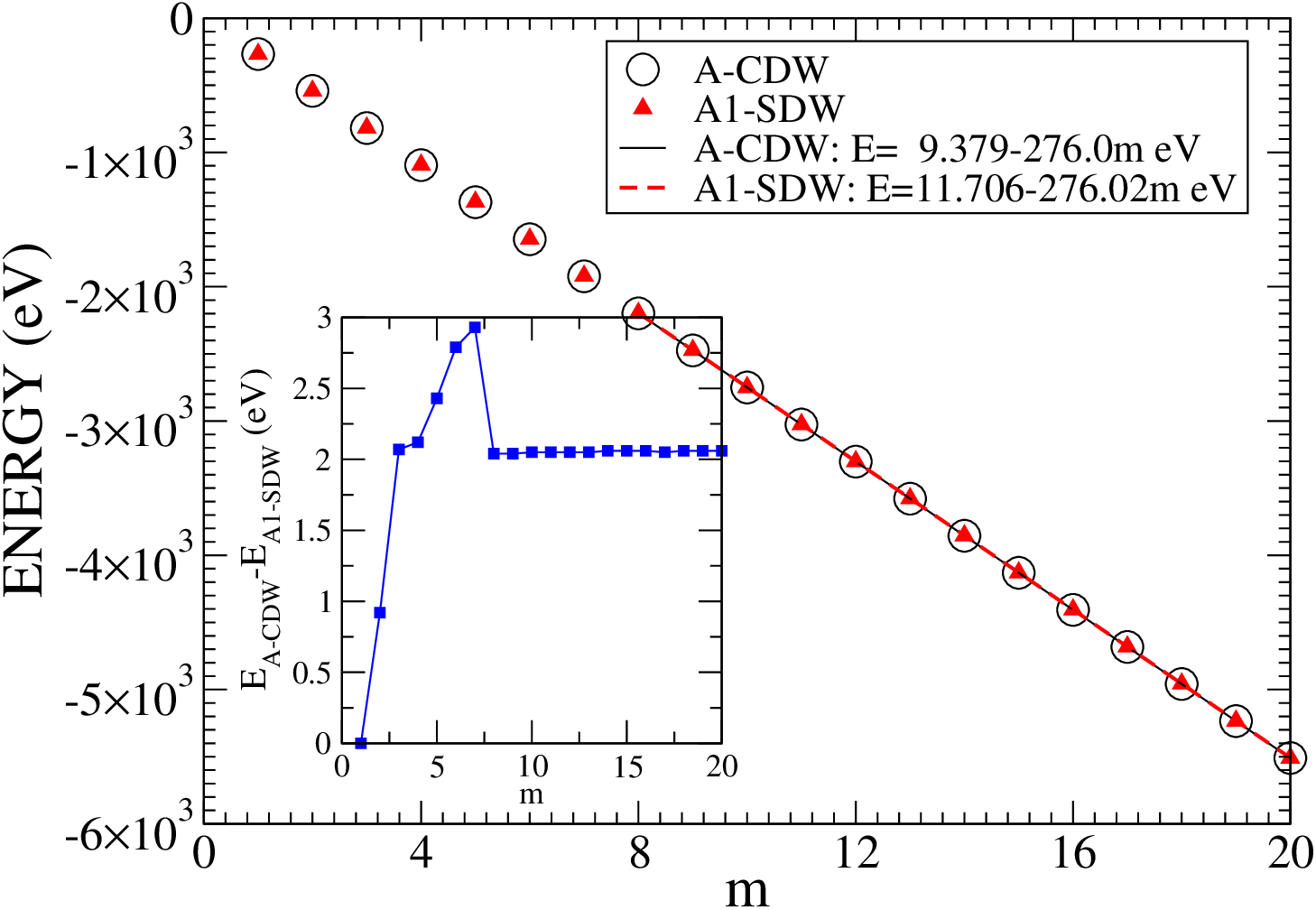}  \\
\caption{(Color online) Energy difference between A-CDW and A1-SDW solutions (see text) in GNR of dimensions $\{1-20,6\}$. For $m \leq 7$ the paramagnetic solution has no CDW.}
\label{ED-SDW-CDW}
\end{figure}

Fig. \ref{ED-SDW-CDW} shows the energy difference between A-CDW and A1-SDW solutions (see text) in GNR of dimensions $\{1-20,6\}$. For $m \leq 7$ the paramagnetic solution has no CDW, whereas the SDW solution shows up for $m > 2$. This explains the irregular behavior of the energy difference at small $m$. Beyond $m=7$, the straight lines fitted to the energies are consistent with the slow decrease of the energy difference as the ribbon length increases (small discrepancies should be ascribed to the fact that while energies vary over thousands of eV,  they differ only in approximately 2 eV, i.e, less than 0.07\%).  
This small energy difference, whose order of magnitude does not vary dramatically with ribbon size,  may justify our conjecture concerning the possibility that many body interactions may favor the CDW solutions.

\section{Concluding Remarks}
Graphene nanoribbons are one of  the best founded hopes for a major jump in the microelectronics (nanoelectronics?) industry.
Initial attempts of fabrication of nanoribbons found many dificulties that hindered obtaining GNR with well-defined shape and size. Very recently, the development of a variety of bottom-up  techniques have led to the reliable production of not only ribbonns with arm-chair edges but also the far more difficult with zig-zag edges. Despite of the good performance of that technique, full control of ribbon shappe and length is not always easy. Avoiding the presence at the ribbon edges of vacancies and kinks is still a major problem. This may be the main cause of discrepancies that still exist amongst  different laboratories. In addition, the large variety of theoretical tools used to tackle the problem not always agree
The main result reported here concerns the different behavior of the energy gap vs ribbon length found for ribbons enlarged either in the armchair or the zig-zag directions. While in the former the gap varies smoothly with the length, it oscillates appreciably in the latter. Unfortunately we do not have any experimental support for this conclusion as most of the experimental studies concern ribbons with length varying in the arm-chair direction. 
 A major unresolved question is whether there is any spin polarisation at the zig-zag edges. All mono-determinantal calculations, the present one included,  indicate that it actually should be. However, as discussed here, charge and spin density waves solutions differ in few eV, while the total ribbon energy soon reaches ten thousand eV, making possible that many-body effects invert that order. As regards experimental verification, there is not yet a trustable strategy to explore the low energy spin physics at graphene nano-ribbons. \\

\begin{acknowledgments}
Financial support by the Spanish "Ministerio de Ciencia e Innovaci\'on  MICINN"
(grants FIS2012-35880 and FIS2015-64222-C2-2-P) and the Universidad de Alicante is gratefully acknowledged.
\end{acknowledgments}


\begin{thebibliography}{99}
\bibitem{WT16}
S. Wang, L. Talirz, C. A. Pignedoli, X. Feng, K. Mu¬llen, R. Fasel and
 P. Ruffieux, Nature Commun., DOI: 10.1038/ncomms11507 (2016).
\bibitem{TR16}
L. Talirz, P. Ruffieux and R. Fasel, Adv. Mater. {\bf 28}, 6222 (2016).
\bibitem{WZ16}
W.-X. Wang, M. Zhou, X. Li, S.-Y. Li, X. Wu, W. Duan, and L. He, Phys. Rev. B 93, 241403(R) (2016).
\bibitem{XL16}
W. Xu and T.-W. Lee. Mater. Horiz. {\bf 3} 186 (2016).
\bibitem{YL16}
C.-N. Yeh,  P.-Y. Lee,  and J.-D. Chai, arXiv:1601.04205v2 [physics.chem-ph] 28 Apr 2016.
\bibitem{WC15}
C.-S. Wu and J.-D. Chai, J. Chem. Theory Comput. {\bf 11}, 2003 (2015).
\bibitem{LG14}
S. Li, C. K. Gan, Y.-W. Son d, Y. P. Feng  S. Y. Quek, Carbon {\bf 76}, 285 ( 2014).
\bibitem{CO13}
Y.C. Chen, D.G. de Oteyza, Z. Pedramrazi, C. Chen, F.R. Fischer and M.F.
Crommie, ACS Nano {\bf 7}, 6123 (2013).
\bibitem{TS13}
L. Talirz, H. Sode, J.M. Cai, P. Ruffieux, S. Blankenburg, R. Jafaar, R. Berger,
X. Feng, K. Mullen, D. Passerone, R. Fasel and C.A. Pignedoli,
J. Am. Chem. Soc. {\bf 135}, 2060 (2013).
\bibitem{RC12}
P. Ruffieux, J.M. Cai, N.C. Plumb, L. Patthey, D. Prezzi, A. Ferretti, E.
Molinari, X.L. Feng, K. Mullen, C.A. Pignedoli and R. Fasel,
ACS Nano {\bf 6}, 6930 (2012).
\bibitem{VC15}
J.A. Verg\'es, G. Chiappe, and E. Louis, Eur. Phys. J. B {\bf 88}, 200 (2015).
\bibitem{KA12}
M. Koch, F. Ample,  C. Joachim, L. Grill,  Nat. Nanotech. {\bf 7}, 713 (2012). 
\bibitem{BP13}
P.B. Bennett, Z. Pedramrazi, A. Madani, Y.C. Chen, D.G. de Oteyza, C. Chen, F.R.
Fischer, M.F. Crommie, and J. Bokor, Appl. Phys. Lett. {\bf 103}, 253114 (2013). 
\bibitem{SC06a}
Y.-W. Son, M. L. Cohen, S. G. Louie, Phys. Rev. Lett. {\bf 97}, 216803 (2006).
\bibitem{SC06b}
Y.-W. Son, M. L. Cohen and  S. G. Louie, Nature {\bf 444}, 347 (2006).
\bibitem{CG09}
A.H. Castro Neto, F. Guinea, N.M.R. Peres, K.S. Novoselov, A.K. Geim,
Rev. Mod. Phys. {\bf 81}, 109 (2009) and references therein.
\bibitem{DA11}
S. Das Sarma, S. Adam, E.H. Hwan and E. Rossi, Rev Mod. Phys. {\bf 83},
407 (2011) and references therein.
\bibitem{LV07}
E. Louis, J.A. Verg\' es, F. Guinea, G. Chiappe, Phys. Rev.  B {\bf 75}, 085440 (2007).
\bibitem{LC14}
Y.Y. Li, M.X. Cen, M. Weinert, L. Li, Nature Commun. {\bf 5}, 4311 (2014).
\bibitem{PP53}
R. Pariser and R.G. Parr, J. Chem. Phys. {\bf 21} 466 (1953).
\bibitem{Po53}
J.A. Pople, Trans. Faraday Soc.  {\bf 49}  1365 (1953).
\bibitem{BC95}
D. Baereswyl, J. Carmelo, D.K. Campbell, F. Guinea and E. Louis, eds., {\it The
Hubbard Model: Its Physics and Mathematical Physics}, NATO ASI Series Vol. 343,
(Plenum Press, New York, 1995).
\bibitem{GS11a}
K. Gundra, A. Shukla, Phys. Rev. {\bf 83},  075413 (2011).
\bibitem{GS11b}
K. Gundra, A. Shukla, Phys. Rev. {\bf 84}, 075442 (2011).
\bibitem{SS10}
P. Sony, A. Shukla, Computer Phys. Commun. {\bf 181}, 821 (2010).
\bibitem{PG12}
P. Potasz, A. D. G{\" u}{\c c}l{\" u}, A. W{\'o}js, and P. Hawrylak,
Phys. Rev. B {\bf 85}, 075431 (2012).
\bibitem{HW12}
M. Hohenadler, S. Wessel, M. Daghofer, and F. F. Assaad,
Phys. Rev. B {\bf 85}, 195115 (2012).
\bibitem{SG11}
E. San-Fabi\'an, A. Guijarro, J.A. Verg\'es, G. Chiappe and E. Louis,
Eur. Phys. J. B {\bf 81}, 253 (2011).
\bibitem{VC09}
J.A. Verg\'es, G. Chiappe, E. Louis, L. Pastor-Abia and E. San-Fabi\'an,
Phys. Rev. B {\bf 79},  094403 (2009).
\bibitem{VS09}
J.A. Verg\'es, E. San-Fabi\'an, L. Pastor-Abia, G. Chiappe and E. Louis,
Phys. Stat. Solidi C {\bf 6}, 2139 (2009).
\bibitem{VS10}
J.A. Verg\'es, E. San-Fabi\'an, G. Chiappe and E. Louis,
Phys. Rev. B {\bf 81}, 085120 (2010).
\bibitem{CL15}
G. Chiappe, E. Louis, E. San-Fabi\'an and J.A. Verg\'es, J. Phys.: Condens. Matter {\bf 27}, 463001 (2015).
\bibitem{Pa86}
D.A. Papaconstantopoulos, {\it Handbook of the Band Structure of Elemental
Solids} (Plenum Press, New York, 1986).
\bibitem{Ma00}
G.D. Mahan, {\it Many-Particle Physics} (Kluwer Academic/Plenum Publishers, New
York, 2000).
\bibitem{Oh64}
K. Ohno, Theor. Chim. Acta {\bf 2}  219 (1964).
\bibitem{MS09}
F. Moscard\'o, E. San-Fabi\'an, Chem. Phys. Lett. {\bf 480}, 26 (2009).
\bibitem{SK84}
K. Seki, U. O. Karlsson, R. Engelhardt, E. E. Koch, W. Schmidt, Chem. Phys. {\bf 91}, 459 (1984).
\bibitem{SS94}
S. W. Staley, J. T. Strnad, J. Phys. Chem. {\bf 98}, 116 (1994).
\bibitem{MD83}
A. Modelli, G. Distefano, D. Jones, Chem. Phys. {\bf 82}, 489 (1983).
\bibitem{NA06}
T. Nakamura, N. Ando, Y. Matsumoto, S. Furuse, M. Mitsui, A. Nakajima, Chem. Lett. {\bf 35}, 888 (2006).
\bibitem{RT02}
J. C. Rienstra-Kiracofe, G. S. Tschumper, H. F. Schaefer III, S. Nandi, G. B. Ellison, Chem. Rev. {\bf 102}, 231 (2002).
\bibitem{note}
It should be noted that the way we increase the size of the ribbon is that followed by the experimental techniques nowadays available \cite{WT16}. This may imply changing the family of the ribbon as defined in Ref. [\onlinecite{YP07}].
\bibitem{NF96}
K. Nakada, M. Fujita, G. Dresselhaus, M.S. Dresselhaus,  Phys. Rev. B {\bf 54}, 17954 (1996).
\bibitem{YP07}
L. Yang, C.-H. Park, Y.-W. Son, M.L. Cohen, S.G. Louie,  Phys. Rev. Lett. {\bf 99}, 186801 (2007).
\end{thebibliography}
\end{document}